\documentstyle[12pt]{article}

\begin{document}
\title{Remote state preparation in higher dimension and the parallelizable
manifold $S^{n-1}$}
\author{Bei Zeng$^1$ and Peng Zhang$^2$ \\
\medskip 1 Department of Physics, Tsinghua University,\\
Beijing,100084,China\\ 2 Institute of Theoretical Physics,
Academia Sinica, Beijing, 100080, China\\
}
\date{February 22, 2001}
\maketitle

\begin{abstract}
This paper proves that the remote state preparation (RSP) scheme
in real Hilbert space can only be implemented when the
dimension of the space is 2,4 or 8. This fact is shown to be
related to the parallelazablity of the $n$-1 dimensional sphere
$S^{n-1}$. When the dimension is 4 and 8 the generalized scheme is
explicitly presented. It is also shown that for a given state with
components having the same norm, RSP can be generalized to
arbitrary dimension case.
\end{abstract}

Remote state preparation (RSP) \cite{1}\cite{2}\cite{3}is called
``teleportation of a known state''. Unlike quantum teleportation \cite{4}%
\cite{5}\cite{6}\cite{7}\cite{8}, in RSP, Alice knows the state which she
will transmit to Bob. Her task is to help Bob to construct a state which is
unknown to him by means of a prior shared entanglement and a classical
communication channel. Recently, Pati has shown that a state of a qubit
chosen from equatorial or polar great circles on the Bloch sphere (i.e. a
state with the components of the same amplitude or with real components) can
be remotely prepared with one cbit from Alice to Bob if they share one ebit
of entanglement[1]. Here, qubit stands for quantum bit whose state is a
superposition of two orthonormal basis $\left| 0\right\rangle $ and $\left|
1\right\rangle $ ; cbit is classical bit carrying classical information;
ebit is the so called entanglement bit usually carrying a Bell state. It is
noted that in Pati's special case, to remotely prepare a state of one qubit,
the entanglement cost is the same as that in teleportation but the classical
information cost is only half of that in teleportation. Most recently, Lo
and Bennett et al have studied the classical information cost for general
state preparation in the scheme of RSP [2][3], using the concepts of
entanglement dilution [9][10], high-entanglement limit and low-entanglement
RSP[3]. They have also investigated the trade-off between entanglement cost
and classical communication cost in RSP [2][3]. However, in Lo or Bennett et
al's protocols, either the entanglement cost or the classical information
cost is more than that in Pati's special case. This fact can be well
understood by considering the geometry of Pati's case: Pati's states lie on
the equatorial or polar great circles on a Bloch sphere. For this reason, we
call the case treated by Pati the ``minimum'' case.

As Pati presents his result only in the qubit case, it is natural to ask
whether his result can be generalized to higher dimension case. It is well
known that as far as teleportation, which transmits an unknown state, is
concerned, the generalization from the qubit case to higher dimension case
is straightforward. In fact, the first $n$-dimensional teleportation protocol
is just given by Bennett et al in their first paper that introduced the
celebrated concept of quantum teleportation [4]. Later the $n$-dimensional
case of teleportation and its mathematical background were studied in more
detail by many other authors [11][12][13][14]. Even in the case concerning
continous variable [15], it can well be tackled [16]. The purpose of this
paper is to seek a generalization of Pati's result to higher dimension
case. It will be shown that one can directly generalize the equatorial case.
On the other hand,the generalization of the polar great circle case is
highly nontrivial.

We first consider the generalization of the polar great circle case (i.e.
the case that the state has real components ). Precisely, we formulate our
problem as follows. Suppose that Alice and Bob can share entangled state
between two identical quantum systems the dimension of the state space of
which is $n.$Choose an orthonormal basis $\left\{ \phi _{i}|i=0,1,\cdots
,n-1\right\} $ of the state space. By measuring the system with respect to a
certain basis, Alice wishes to prepare a quantum state of the form
\[
\left| \Psi \right\rangle =\sum_{i=0}^{n-1}a_{i}\left| \phi
_{i}\right\rangle
\]%
at Bob, where the coefficients are real numbers. Between Alice and Bob there
is a classical channel capable of transmitting information carried by a
``classical bit'' that can take $n$ different values, say, $0,1,\cdots .n-1$%
. By prior agreement, each value carried by the ``classical bit'' can be
corresponded to a unitary operation on the quantum system at Bob. That is to
say, when Bob receives a value $i$ he will exert a certain unitary operation
$U_{i}$ on his system. Now our question is: for the above minimum RSP procedure
to
be realizable what condition should the dimension $n$ satisfy? By
convention,in the procedure of RSP the maximally entangled state shared by
Alice and Bob, will be the EPR state
\[
\left| \Phi \right\rangle _{AB}=\frac{1}{\sqrt{n}}\left(
\sum_{i=0}^{n-1}\left| \phi _{i}\right\rangle \otimes \left| \phi
_{i}\right\rangle \right)
\]

Remark. In reference [1], the EPR state is$\left| \Phi \right\rangle _{AB}=%
\frac{1}{\sqrt{2}}$ $\left( \left| 0\right\rangle \otimes \left|
1\right\rangle -\left| 1\right\rangle \otimes \left| 0\right\rangle \right) $%
, which is a little different from the EPR state we use here. But there is
no essential difference.

To prepare the state $\left| \Psi \right\rangle $ in a remote place, similar
to the Pati's protocol in qubit case, Alice needs to find a set of
orthonormal basis \{$\left| \Psi _{i}\right\rangle $\}$_{i=0}^{n-1}$ with
respect to which the measurement is done on her system. The EPR state $%
\left| \Phi \right\rangle _{AB}$ can be written as $\left| \Phi
\right\rangle _{AB}=\frac{1}{\sqrt{n}}\sum_{i}\left| \Psi _{i}\right\rangle
\otimes \left| \Omega _{i}\right\rangle $ . Here $\left| \Omega
_{i}\right\rangle =\sum_{j,k}\left| \Psi _{j}\right\rangle \ast \lbrack
\left\langle \Psi _{j}\mid \phi _{k}\right\rangle \left\langle \Psi _{i}\mid
\phi _{k}\right\rangle ],$ $i=0,...,n-1$. We notice that \{$\left| \Omega
_{i}\right\rangle \}_{i=0}^{n-1}$is a set of orthonormal vectors. To realize
the minimum RSP task, there should exist $n$ unitary operators $U_{i}$ $%
(i=0,1,\cdots ,n-1)$ independent of $\left| \Psi \right\rangle $ such that $%
\left| \Omega _{i}\right\rangle =U_{i}\left| \Psi \right\rangle $. If such
unitary operators do exist, then Alice can measure her system with respect
to the basis \{$\left| \Psi _{i}\right\rangle $\}$_{i=0}^{n-1}$ and get a
state $\left| \Psi _{i}\right\rangle $. Then through the classical
communication channel, she can send Bob the value $i.$ After receiving the
message, Bob will be able to construct the target state $\left| \Psi
\right\rangle $ by letting his system experience the unitary evolution $%
U_{i}$, according to their prior agreement. It turns out that the
requirement that such unitary operators $U_{i}$'s exist imposes very strong
restriction on the dimension of the state space. Before proceeding along
with the discussion, let us prepare some terminology about parallelizable
manifold.

Let $M$ be a manifold of dimension $n$. The tangent space $T_{x}M$ is well
defined for every point $x\in M$. A continuous vector field $v$ in $M$ is a
continuous function which assigns a vector $v(x)\in T_{x}M$ to every $x\in M$%
. By a $k-$field we mean a $k-$tuple $v_{1},v_{2},\cdots ,v_{k}$ of
continuous vector fields on $M$, such that the vectors $v_{1}(x),\cdots
,v_{n}(x)$ at each point $x\in M$ are linearly independent. The largest $k$
for which a $k-$field exists is called Span($M$). If Span($M$)$=n$, then the
manifold is said to be parallelizable. It is a difficult problem to
determine Span($M$) for any given manifold. But we have the following deep
result[17].

Theorem. The sphere $S^{n-1}$ is parallelizable only for $n=1,2,4,8$.

We proceed to prove the following interesting result.

Proposition. If the minimum RSP scheme is realizable in $n$-dimensional real Hil
bert space, then the
sphere $S^{n-1}$ is parallelizable.

Proof. From the above discussion, if RSP is realizable there should exist $n$
unitary operators $U_{i}(i=0,1,\cdots ,n-1)$ such that $\left| \Omega
_{i}\right\rangle =U_{i}\left| \Psi \right\rangle $. As pointed out above, $%
\left\{ \left| \Omega _{i}\right\rangle |i=0,1,\cdots ,n-1\right\} $ is a
set of orthonormal vectors. Thus we have%
\[
\left\langle \Psi |U_{0}^{-\dagger }U_{i}|\Psi \right\rangle =0,i=1,2,\cdots
,n-1
\]%
Write $U_{0}^{-\dagger }U_{i}$ as
\[
U_{0}^{-\dagger }U_{i}=V_{i}+\sqrt{-1}W_{i}
\]%
where $V_{i}$ and $W_{i}$ are real matrices. Then it follows that
\[
\left\langle \Psi |V_{i}|\Psi \right\rangle =0,i=1,2,\cdots ,n-1
\]%
as $\left| \Psi \right\rangle $ has real coefficients. If we only consider the c
ase
 $W_i=0$, i.e. the minimum RSP scheme in real Hilbert space, this $\{V_i|\Psi\rangle\}$
 ($i=0,1,2,...n-1$)form an orthonormal basis of $n$-dimensional real Hilbert spa
ce.
 Obviously, $\left|
\Psi \right\rangle $ can be regarded as a point on $S^{n-1}.$ Thus the map $%
\left| \Psi \right\rangle \rightarrow V_{i}\left| \Psi \right\rangle $
defines an $(n-1)-$ field on the manifold $S^{n-1}$. This means that $%
S^{n-1} $ is parallelizable.

Now we are prepared to present the main result of this paper.

Main Theorem. Minimum RSP is realizable in real Hilbert space
if and only if the dimension of the space is $1,2,4$ or $8$.

We notice that the ``only if'' part of the theorem is a direct consequence
of the above proposition and the cited theorem preceding it. To prove the
``if part'' of the theorem we only need to show that when $n=1,2,4$ or $8$
there exist real unitary matrices $V_{i}(i=0,1,\cdots ,n-1)$ such that for
any $\Psi $ with real coefficients $\left\{ V_{i}\left| \Psi \right\rangle
|i=0,1,\cdots ,n-1\right\} $ is a orthonormal basis of the state space.
Indeed if such unitary matrices exist then the EPR state can be rewritten as
\[
\left| \Phi \right\rangle _{AB}=\frac{1}{\sqrt{n}}\left(
\sum_{i=0}^{n-1}\left| \Psi _{i}\right\rangle \otimes \left| \Psi
_{i}\right\rangle \right)
\]%
where $\left| \Psi _{i}\right\rangle =V_{i}\left| \Psi \right\rangle .$Then
it is clear that RSP can be realized. Since the 1-dimensional case is
trivial and the 2-dimensional case have been dealt with by Pati [1], in the
following we only consider the cases of $n=4$ and $n=8$.

We observe that the existence of the above mentioned $V_{i}$ is closely
related to the existence of $(n-1)-$field on the manifold $S^{n-1}.$So at
this point it is enlightening to recall the marvelous method of relating
the dimension $n$ of a division algebra over the real number field $R$ to
the parallelizability of the manifold $S^{n-1}.$It turns out that by this
method we can find the $V_{i}$'s we need.

It is noticed that if $A$ is a division algebra of dimension $n$, one can choose a vector space
isomorphism to $A$ onto $R^{n}$ and transfer the multiplication defined on $%
A $  to $R^{n}$\cite{17}. Let $e_{1},e_{2},\cdots ,e_{n}$ be the standard basis
vectors of $R^{n}$ and let $y\in S^{n-1}$. Then the vectors $e_{1}\bullet
y,e_{2}\bullet y,\cdots ,e_{n}\bullet y$ are linear independent. If we
orthonormalize them we obtain $n$ vectors $V_{0}(y),V_{1}(y),\cdots
,V_{n-1}(y)$. The vectors $V_{1}(y),\cdots ,V_{n-1}(y)$ are tangential to $%
S^{n-1}$ at the point $V_{0}(y)$.They define an $(n-1)-$field on $S^{n-1}$.
Now it is not difficult to see that when we take $A$ to be the quarternion
algebra and the ontonion algebra, whose dimension is $4$ and $8$
respectively, these $V_{i}$'s are exactly what we need. This finishes the
proof of the main theorem.

To illustrate the above procedure we explicitly calculate the $V_{i}$'s as
follows.

When n=4, we consider the quaternion field H. A quaternion in H can be
expressed as$A=a_0e_0+a_1e_1+a_2e_2+a_3e_3$ , where $\left\{ e_i\right\}
_{i=0}^3$form the standard basis of quaternion. According to the rules of
Hamilton multiplication [17], two quaternions' Hamilton multiplication can
be calculated as follows
\begin{eqnarray*}
A\bullet B &=&(a_0e_0+a_1e_1+a_2e_2+a_3e_3)\bullet \left(
b_0e_0+b_1e_1+b_2e_2+b_3e_3\right) \\
&=&\left( a_0b_0-a_1b_1-a_2b_2-a_3b_3\right) e_0+\left(
a_0b_1+a_1b_0+a_2b_3-a_3b_2\right) e_1 \\
&&+\left( a_0b_2-a_1b_3+a_2b_0+a_3b_1\right) e_2+\left(
a_0b_3+a_1b_2-a_2b_1+a_3b_0\right) e_3
\end{eqnarray*}

Of course, with the usual addition and scalar product , H can be considered
as a vector space over R, which is isomorphic to R$^{4}$ and $e_{0}$, $e_{1}$%
, $e_{2}$, $e_{3}$ form a set of natural basis of this linear space. The
inner product in H can be defined as$\left\langle e_{i},e_{j}\right\rangle
=\delta _{ij}$ , i.e.$\left\langle A,B\right\rangle
=\sum_{i=0}^{3}a_{i}b_{i} $ . For an arbitrary unit vector A which satisfies
$\left\langle A,A\right\rangle =1$, we can define a set of vectors$%
\{A_{i}=e_{i}\bullet A\}_{i=0}^{3}$ . Using the property of division algebra
[17], we have $\left\langle A_{i},A_{j}\right\rangle =\left\langle
e_{j},e_{i}\right\rangle \left\langle A,A\right\rangle =\left\langle
e_{j},e_{i}\right\rangle =\delta _{ij}$. Therefore,$\left\{ A_{i}\right\}
_{i=0}^{3}$ is a set of orthonormal basis. It is easy to see that $A_{0}=A$
and the orthonormal transformations $\left\{ V_{i}\right\} _{i=0}^{3}$ that
transform $A$ to $\left\{ A_{i}\right\} _{i=0}^{3}$ are independent of $A$ .
Therefore, $\left\{ V_{i}\right\} _{i=0}^{3}$ are just what we want to find.

A direct calculation following the above steps gives the following
result in the 4-dimension case:

$V_0=I,$ $V_1=\left[
\begin{array}{cc}
-i\sigma _y & 0 \\
0 & -i\sigma _y
\end{array}
\right]$

$V_2=\left[
\begin{array}{cc}
0 & -\sigma _z \\
\sigma _z & 0%
\end{array}
\right],V_3=\left[
\begin{array}{cc}
0 & -\sigma _x \\
\sigma _x & 0%
\end{array}
\right]$

When $n=8$, using the rules of Cayley multiplication [17], similarly we can
get $\left\{ V_{i}\right\} _{i=0}^{7}$ . The result is as follows.

$V_0=I,V_1=\left[
\begin{array}{cccc}
-i\sigma_y & 0 & 0 & 0 \\
0 & -i\sigma_y & 0 & 0\\
0 & 0 & -i\sigma_y & 0 \\
0 & 0 & 0 & -i\sigma_y %
\end{array}
\right]$

$V_2=\left[
\begin{array}{cccc}
0 & -\sigma_z & 0 & 0 \\
\sigma_z & 0 & 0 & 0 \\
0 & 0 & 0 & -I \\
0 & 0 & I & 0%
\end{array}
\right],V_3=\left[
\begin{array}{cccc}
0 & -\sigma_x & 0 & 0 \\
\sigma_x & 0 & 0 & 0 \\
0 & 0 & 0 & -i\sigma_y \\
0 & 0 & -i\sigma_y & 0%
\end{array}
\right]$

 $V_4=\left[
\begin{array}{cccc}
0 & 0 & -\sigma_z & 0 \\
0 & 0 & 0 & I \\
\sigma _z & 0 & 0 & 0 \\
0 & -I & 0 & 0%
\end{array}
\right] ,V_5=\left[
\begin{array}{cccc}
0 & 0 & -\sigma _x & 0 \\
0 & 0 & 0 & i\sigma _y \\
\sigma _x & 0 & 0 & 0 \\
0 & i\sigma _y & 0 & 0%
\end{array}
\right] ,$

$V_6=\left[
\begin{array}{cccc}
0 & 0 & 0 & -I \\
0 & 0 & -\sigma _z & 0 \\
0 & \sigma _z & 0 & 0 \\
I & 0 & 0 & 0%
\end{array}
\right] ,V_7=\left[
\begin{array}{cccc}
0 & 0 & 0 & -i\sigma _y \\
0 & 0 & -\sigma _x & 0 \\
0 & \sigma _x & 0 & 0 \\
-i\sigma _y & 0 & 0 & 0%
\end{array}
\right] $
\\
Moreover, in general the case that $W_{i}\neq 0$, i.e. minimum
RSP scheme for states with real components in complex Hilbert space should
be taken into account. We conjecture that even in this case, minimum RSP scheme
can only be implemented when the dimension of the space is $2,4$ or $8$.
\\
Now we consider the generalization of RSP scheme of the equatorial case. In
this case, the state to be remotely prepared can be written in the form
\[
\left| \Psi \right\rangle =\sum_{\alpha =0}^{n-1}\frac{1}{\sqrt{n}}%
e^{i\theta _{\alpha }}\left| \alpha \right\rangle
\]

Without loss of generality, we set $\theta _{0}=0$ . We will show
the RSP scheme for such states is realizable whatever the dimension $n$ is.
It is easily seen that $\left\{ \left| \Psi _{\alpha }\right\rangle \mid
\left| \Psi _{\alpha }\right\rangle =\frac{1}{\sqrt{n}}\sum_{\beta
=0}^{n-1}e^{\frac{2\pi i}{n}\alpha \beta }e^{i\theta _{\alpha }}\left| \beta
\right\rangle \right\} _{\alpha =0}^{n-1}$ \ is an orthonormal basis in the
n-dimensional case, and that the unitary transformation $U_{\alpha }:$ $%
U_{\alpha }\mid U_{\alpha }\left| \Psi \right\rangle =\left| \Psi _{\alpha
}\right\rangle $ is independent of $\left| \Psi \right\rangle $ . As the
first step to remotely prepare $\left| \Psi \right\rangle $ , Alice needs to
do a local unitary transformation $U_{A}(\left| \Psi \right\rangle )$ on her
particle. Here, $U_{A}(\left| \Psi \right\rangle )$ is defined as
\[
U_{A}(\left| \Psi \right\rangle )=\sum_{\alpha =1}^{n-1}\left| \alpha
\right\rangle \left\langle n-\alpha \right| e^{i\left( \theta _{\alpha
}+\theta _{n-\alpha }\right) }+\left| 0\right\rangle \left\langle 0\right|
\]

Thus we have
\[
U_A(\left| \Psi \right\rangle )\otimes I_B\left| \Phi \right\rangle
_{AB}=\sum_{\alpha =0}^{n-1}\frac 1{\sqrt{n}}\left| \Psi _\alpha
\right\rangle \otimes \left| \Psi _\alpha \right\rangle
\]

Here $\left| \Phi \right\rangle _{AB}$ is the EPR state $\left| \Phi
\right\rangle _{AB}=\sum_{\alpha =0}^{n-1}\frac{1}{\sqrt{n}}\left| \alpha
\right\rangle \otimes \left| \alpha \right\rangle $ of the entangled pair
which is prior shared by Alice and Bob. After the transformation, Alice can
measure her particle with respect to the basis \{$\left| \Psi _{\alpha
}\right\rangle $\}$_{\alpha =0}^{n-1}$ and tell her result to Bob. Then Bob
can do the unitary transformation $U_{\alpha }^{-1}$ to get the state $%
\left| \Psi \right\rangle $ . This implements the RSP task.

In summary, this paper generalizes Pati's minimum RSP scheme to the case of
higher dimension. We have shown that the minimum RSP scheme in real Hilbert
space can be implemented only when the dimension is $2,4$ or $%
8$, while the equatorial case can be generalized without restriction on the
dimension.However, whether the minimum RSP scheme for the states
with real components in complex Hilbert space is realizable in other dimensional
space rather than $2,4$ and $8$ dimension needs further investigation.

The authors gratefully thank Professors C. P. Sun for initial discussion on RSP.
The authors are also grateful to Dr. X. N. Wu ,
Dr. Y. Q. Li and especially Professor X. F. Liu for discussions with them. This work is partially supported by the
NSF of China.

\end{document}